\documentclass[aps,prl,twocolumn,footinbib,floatfix]{revtex4-1}

\usepackage{CJK}
\usepackage{amsmath}
\usepackage{amsfonts}
\usepackage{bm}
\usepackage{soul}
\usepackage[caption=false,subrefformat=parens,labelformat=parens]{subfig}
\usepackage{color} 
\usepackage[table,dvipsnames]{xcolor}
\usepackage[colorlinks,linktocpage,bookmarks=false]{hyperref}

\newcommand\myshade{85}
\definecolor{myrulecolor}{RGB}{150,20,0}
\colorlet{mylinkcolor}{violet}
\colorlet{mycitecolor}{YellowOrange}
\colorlet{myurlcolor}{Aquamarine}
\hypersetup{
	linkcolor  = myurlcolor!\myshade!black,
	citecolor  = mycitecolor!\myshade!black,
	urlcolor   =  myrulecolor!\myshade!black,
}

\usepackage{todonotes}

\usepackage{graphicx}
\graphicspath{{FIG/}}
\usepackage{multirow}
\usepackage{braket}

\newcommand{\beq}{\begin{equation}}
\newcommand{\eeq}{\end{equation}}
\newcommand{\bea}{\begin{eqnarray}}
\newcommand{\eea}{\end{eqnarray}}

\newcommand{\Tr}{\text{Tr}}

\renewcommand\[{\begin{equation}}
\renewcommand\]{\end{equation}}

\makeindex

\begin{document} 
	\begin{CJK*}{UTF8}{gbsn} 
		\title{Geodesic string condensation from symmetric tensor gauge theory: \\a unifying framework of holographic toy models}
		\author{Han Yan (闫寒)}
		\email{han.yan@oist.jp}
		\affiliation{Theory of Quantum Matter Unit, Okinawa Institute of Science and Technology Graduate University, Onna-son, Okinawa 904-0495, Japan}
		\date{\today}
\begin{abstract}
In this work we reason that 
there is a universal picture for several different holographic toy model constructions,
and  a gravity-like bulk field theory that gives rise it.
First, we observe that
the perfect tensor-networks and hyperbolic  fracton models
are both equivalent to the even distribution of 
bit-threads on geodesics in the AdS space. 
Such picture is also a natural ``leading-order'' approximation 
to the holographic entanglement properties.
Then,
we argue that
the  rank-2 U(1)  theory with linearized diffeomorphism as its gauge symmetry,
also known as a case of Lifshitz gravity,
is the bulk field theory behind such picture.
The Gauss' laws and spatial curvature require the
electric field lines along the geodesics
to be the fundamental dynamical variables,
which lead to geodesic string condensation.
These results provide an intuitive way to understand the
entanglement structure of gravity in AdS/CFT.
\end{abstract}
\maketitle
\end{CJK*}


\section{Introduction}
Modern physics has entered an exciting era of intertwined influences between quantum many-body systems, quantum gravity, and quantum information. 
Many of these interdisciplinary conversations
revolve around 
the holographic principle \cite{Hooft1974,Susskind1995} and anti-de Sitter/conformal field theory (AdS/CFT) correspondence \cite{Maldacena1999,Witten1998}.
As a conjectured duality between quantum gravity in $(d+1)$-dimensional asymptotically AdS spacetime
and a $d$-dimensional CFT on its boundary,
the holographic principle
is a profound insight of quantum gravity,
and also acts as a powerful tool for condensed matter problems \cite{Hartnoll}.

In 2006,
Ryu and Takayanagi conjectured that the entanglement entropy of
a boundary CFT segment is measured by the area of the corresponding extremal covering surface 
in the AdS geometry \cite{Ryu2006,Ryu2006a}.
This conjecture reveals the intimate relation between entanglement
and geometry in quantum gravity.
Following the insight of 
Swingle \cite{Swingle2012},
various tensor-network holographic toy models were built
\cite{Swingle2012,Pastawski2015,Almheiri2015,Yang2016,Hayden2016,Qi2018,Harlow2017CMaP},
and they uncover the quantum-informational correcting feature of holographic entanglement.

On the condensed matter hemisphere,
the fracton states of matter were 
studied intensively in recent years
 \cite{
ChamonPhysRevLett.94.040402,YoshidaPhysRevB.88.125122,BRAVYI2011839,Haah2011,Vijay2015,Vijay2016,Pretko2017a,Pretko2017b,Nandkishoreannurev}.
The gapless versions of fracton states, namely rank-2 U(1) [R2-U1] gauge theories
\cite{XuPRB06,rasmussen2016stable,Pretko2017a,Pretko2017b},
were found to be different cases of Lifshitz gravity \cite{Horava2009PhysRevD,Xu2010PhysRevD}.
The charge excitations dubbed ``fractons'' also show gravitational attractions \cite{Pretko2017PhysRevD}.
Inspired by these discoveries, 
a toy fracton model in AdS space was
studied and shown to satisfy holographic properties
in a  similar fashion as the holographic tensor-networks \cite{Yan2018arXiv,yan2019hyperbolic}.

Some important questions remain unanswered despite these progress.
First, what is the connection between the different holographic toy models?
Is there a universal picture behind them?
Furthermore, how can we derive its continuous limit from a  bulk field theory,
and how is the bulk theory related to gravity?
This is in particular intriguing for the perfect tensor models, 
since they are clever constructions directly based on holographic entanglement properties,
but their correspondence to any concrete bulk field theory is still unknown.
Although  the hyperbolic fracton model is a spin model in the bulk,
it is still far from a  field theory that shows satisfactory resemblance to gravity.

In this work,
we
advance our understanding of holographic toy models
 by addressing these two questions.
First, we point out that there is a universal picture behind 
 different constructions
of holographic toy models:
a homogeneous and isotropic distribution of bit-threads (or up to some lattice discretization).
We then show that the traceful, vector charged R2-U1, 
a theory with linearized diffeomorphsim as its gauge symmetry,
gives rise to this continuous bit-thread picture.
We reason that
in the presence of spatial curvature,
the gauge symmetry, and the consequent Gauss' laws,
only allow electric field lines along a geodesics
to be the fundamental dynamical variables (magnetic field),
Any loop configurations like in conventional U(1) theory
are forbidden.
Hence the entanglement structure 
is determined by the ``geodesic string condensation,''
which is exactly the continuous bit-thread picture.
As such, we establish
the connection 
between the holographic toy models
and a concrete gravity-like bulk field theory,
and shows how entanglement structure emerges from
linearized diffeomorphism.

\section{Bit-thread type holographic toy models as a universal picture} \label{SEC_II_bitthread_model}

\begin{figure}[h]
	\centering
	\includegraphics[width=0.3\textwidth]{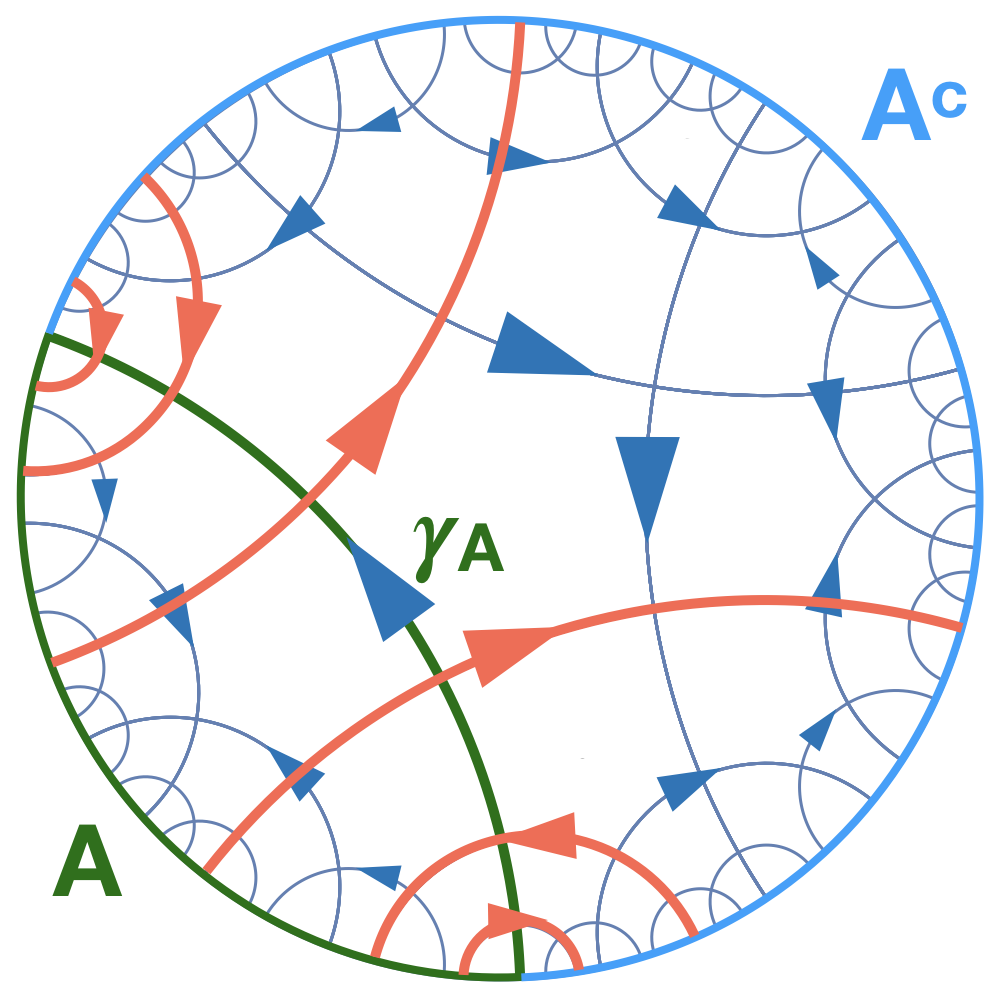}
	\caption{
		Universal picture of holographic toy models: 
		bit-threads distributed evenly on the hyperbolic lattice.
		In the continuous case it is bit-threads distributed homogeneously
		and isotropically in AdS space.
		The bit-threads connecting boundary subregion $A$ and its complement $A^c$
		are highlighted in orange.
		Their number
		is proportional to the length of covering geodesic $\gamma_A$,
		which yields the Ryu-Takayanagi formula (Eq.~\eqref{EQN_RT_formula}).
	} 
	\label{Fig_Bit_Thread}
\end{figure}

This work is motivated by recent progress
on toy models of holography.
They are based on two different constructions:
the hyperbolic fracton model \cite{Yan2018arXiv,yan2019hyperbolic}
and
the perfect tensor-networks
\cite{Swingle2012,Pastawski2015,Almheiri2015,Yang2016,Hayden2016,Qi2018,Harlow2017CMaP}.
But we observe that they belong to a universal picture:
They are equivalent to 
bit-threads arranged on a tessellation
of the hyperbolic disk.
A bit-thread is a line
with entangled qubits (or more generally, 
we can consider any quantum/classical degrees of freedom)
at its two ends
 \cite{Freedman2017CMaPh,Headrick2018CQG,Chen2018arXiv}.
A flow of the bit-threads in the AdS space,
when saturating 
minimal covering surface of a boundary subregion,
gives the correct entanglement entropy 
described by the Ryu-Takayanagi formula [RT-formula]
\cite{Ryu2006,Ryu2006a}
\[
\label{EQN_RT_formula}
S_\text{A} = \frac{\text{area}(\gamma_A)}{4G_N}		.
\]

The hyperbolic fracton model
is dual to the eight-vertex model defined on the edges of the pentagon tessellation (Fig.~\ref{Fig_Bit_Thread})
\cite{yan2019hyperbolic}.
At low temperature,
the eight-vertex model becomes 
a web of independent one-dimensional chains with ferromagnetic couplings.
Each chain is then a classical bit-thread with its two ends correlated. 

Recent work by  Jahn et al. \cite{Jahneaaw0092,alex2019majorana} 
show that
the perfect tensor-network can be described by 
Majorana modes via Jordan-Wigner transformation.
In the Mjaorana fermion language,
the tensor-network state becomes
a collection of Majorana dimers.
Each dimer locates at the two ends of a geodesic on the tensor tessellation,
which is the bit-thread.

Following these observations, a universal picture of the holographic toy model emerges:
by arranging the bit-threads in the AdS space
 homogeneously and isotropically in the continuous limit,
or on a regular tessellation in the discrete case, 
the simplest toy model of holography can be constructed. 

Such toy models capture the RT-formula
for entanglement entropy of any \textit{connected}
boundary subregion.
Instead of adjusting the bit-thread flow to saturate the target covering surface
like in the original proposal \cite{Freedman2017CMaPh},
the bit-threads in this picture are in a fixed configuration,
but the RT-formula is satisfied due to their even distribution in the bulk.
The bulk information is defined in the dual model 
in both the hyperbolic fracton model
and the perfect tensor-networks.
Its reconstruction obeys the Rindler reconstruction rule \cite{Raamsdonk2009arXiv,Dong2016PRL},
again when the boundary subregion is connected.

It can be viewed as a ``leading order''
approximation of the entanglement structure of AdS/CFT.
Built upon a collection of two-body entangled qubits/bits
only,
it naturally
fails to capture the finer entanglement structure of genuine 
gravitational AdS/CFT.

For example,
the entanglement spectrum of a boundary subregion
is always flat, thus the $n$th-R\'{e}nyi entanglement entropy
\[
S_n(\rho_A) = \frac{1}{1-n}\log \Tr \rho_A^n
\]
has no $n$-dependence,
while in AdS/CFT the  $n$-dependence is non-trivial \cite{Dong2016NatComm}. 
Also,
such models   deviate from the RT-formula
when the boundary subregion has multiple disconnected 
components.
This deviation is due to the bit-threads connecting different components 
of the boundary subregion,
which is discussed in Ref.~\cite{yan2019hyperbolic}.

In Table.~\ref{TAB_compare} , we have summarized the comparison
between genuine AdS/CFT, the bit-thread type toy models,
and for completeness also the holographic random  tensor-networks proposed by Yang  et al.
\cite{Yang2016,Hayden2016,Qi2018}.
The random tensor-network satisfies RT-formula for arbitrary boundary subregion,
and does not belong to the universal picture proposed here.

\begin{table*}
\begin{tabular}{c c c c}
\toprule
	\quad &  \quad  AdS/CFT \quad &\quad Bit-thread type toy models  
	& \quad random tensor-networks
	\\ 
	\colrule
	RT-formula for connected boundary subregion & Yes & Yes & Yes \\ 
	 
	RT-formula for disconnected boundary subregion & Yes & No & Yes \\ 
	 
	$n-$dependence of R\'{e}nyi entropy & Yes & No & No  \\ 
	 
	Non-flat entanglement spectrum & Yes & No& No \\ 
\botrule
\end{tabular} 
\caption{
Comparison of the holographic entanglement properties
between genuine AdS/CFT, bit-thread type holographic toy models,
and random tensor-networks.
The bit-thread type holographic toy models,
as a ``leading order'' approximation to holographic entanglement entropy,
capture the Ryu-Takayanagi formula (Eq.~\eqref{EQN_RT_formula}) for 
connected boundary subregion, but not other finer details.
\label{TAB_compare}
}
\end{table*}

These observations lead to the question
this work  addresses:
what is the bulk field theory
that gives rise to the bit-thread type 
of holographic toy models?
These toy models capture the RT-formula and Rindler reconstruction
at ``leading order,'' but fails at ``higher order,'' or the finer entanglement structure.
Thus, 
the reasonable  speculation is that
such bulk theory cannot be the full-fledged
general relativity,
but it has to share certain essential features of gravity
or is a special limit/case of it.

We will show that, 
the bulk theory that describes the bit-thread type toy models 
is the Lifshitz gravity \cite{Horava2009PhysRevD,Sotiriou_2011} in the high energy theory literature, 
or the traceful, vector charged rank-2 U(1) gauge theory \cite{XuPRB06,Xu2010PhysRevD,rasmussen2016stable,Pretko2017a,Pretko2017b,Slagle2018} in the condensed matter physics literature.
As a special case of general relativity,
its gauge symmetry is the spatial part of the linearized diffeomorphism.
As we will elaborate,
the consequence of such gauge symmetry
is that the electric field lines can only travel along 
the geodesics in AdS space, instead of forming local loops like in conventional
gauge theory.
Hence,
the entanglement structure is the continuous bit-thread distribution.

\section{Rank-2 U(1) Theory and Its Flat-Space Dynamics}\label{SEC_III_lefshitz_gravity}

Let us first quickly review 
the traceful, vector-charged version of R2-U1 theory  \cite{XuPRB06,rasmussen2016stable,Pretko2017a,Pretko2017b}.
Here we work in two-dimensional space, but the physics 
naturally extends to higher dimensions.

The R2-U1 theory has gauge symmetry
\[
\label{EQN_Gauge_Trans}
	A^{ij} \rightarrow A^{ij} + \partial^i\lambda^j +  \partial^j\lambda^i 		.
\]
Taking $A^{ij}$ as the perturbation of the metric $A^{ij} = h^{ij} - \delta^{ij}$ in general relativity,
the  transformation  is the linearized limit of diffeomorphism.

The gauge symmetry corresponds to the 
Gauss' laws of the electric field at low energy.
In this case,
the electric field is a rank-2 symmetric tensor
\[
\label{EQN_R2U1_G1}
E^{ij}  =  E^{ji}		.
\]
The Gauss' laws imposed on the electric field is
\[
\label{EQN_R2U1_G2}
\partial_{i}E^{ij} = 0	.
\]
We take both Eq.\eqref{EQN_R2U1_G1} and \eqref{EQN_R2U1_G2} to be the Gauss' laws at the low energy sector 
of the theory.

The charge for such diffeomorphism-like gauge theory is a vector, defined as
\[
\rho^i = \partial_j E^{ij}		.
\]
Beside the total vector charge conservation,
the symmetric condition imposes an additional conservation law
\[
[\int \mathop{dv} \bm{\rho} \times {\bf x} ]^k= \int \mathop{dv} \epsilon^k_{\ ij} x^i \partial_l E^{jl} = - \int \mathop{dv}
\epsilon^k_{\ ij}
 E^{ij}= 0.
\]
This restricts the movement of 
a vector charge $\bm{\rho}$.
The charge $\bm{\rho}$ can only move in the direction of  itself.
It has crucial consequences in the entanglement structure,
as we shall see.

Finally, in the \textit{flat} space, 
the magnetic field is the simplest gauge symmetry-invariant term,
\[
\label{EQN_R2U1_B}
B = \epsilon^{ai} \epsilon^{bj} \nabla_a \nabla_b A_{ij}		.
\]
And the Hamiltonian is 
\[
\label{EQN_R2U1_Ham}
\mathcal{H}_\text{R2-U1-flat} =U  E_{ij} E^{ij} + t B^2		.
\]
The dynamics  $B$, however,
do not survive in the presence of spatial curvature,
as we will explain later.

The Hamiltonian of Eq.~\eqref{EQN_R2U1_Ham}
is also a case of Lifshitz gravity \cite{Xu2010PhysRevD}.
Treating $A^{ij}$ as the perturbation of the metric,
the magnetic field squared term $B^2$ is equivalent to $R^2$, $R$ being the Ricci scalar.
Here, the  conventional linear term $R$ and cosmological constant $\Lambda$ in general relativity,
as well as the self-interacting, non-linear terms are forbidden  due to the time-reversal,
lattice translation, and spatial reflection symmetries. 
This was carefully analyzed in Ref.~\cite{Xu2010PhysRevD}.
So the theory of Eq.~\ref{EQN_R2U1_Ham} can be viewed as a special version of linearized gravity.

\section{Entanglement  structure from  Gauge symmetry: conventional U(1) as an example}\label{SEC_IV_entanglement}

Before examining the entanglement structure of the R2-U1 in AdS space,
let us first review the topological entanglement entropy in the conventional U(1) gauge theory
from the condensed matter point-of-view \cite{Levin2005PhysRevB,Levin2006PhysRevLett,Kitaev2006PhysRevLett,Pretko2016PhysRevB}.
It is the string-net condensation picture proposed by Levin and Wen \cite{Levin2005PhysRevB}.
This helps to understand the logical chain of how gauge symmetry determines the 
entanglement structure.
The gauge symmetry for conventional U(1) theory
\[
A^i \rightarrow A^i + \partial^i \lambda
\]
as our starting point 
determines
the  Gauss' law to be  electric charge conservation
\[
\partial_iE^i  = 0 
\]
At low energy,
the operations of gauge field $A^x,\ A^y$  are
to construct  microscopic  electric fields,
or dipoles (cf. Table.~\ref{TAB_BIG}).
The gauge field operators respect the charge conservation
globally, but not locally.
Mathematically, that is to say the gauge field themselves
are not gauge invariant.

To respect the Gauss' law in any infinitesimal, local subregion,
the dipoles operators have to be connected head-to-tail together
to form a loop.
The minimal loop is the magnetic field ${B} = \epsilon^{ij}\nabla_i A_j$ (cf. Table.~\ref{TAB_BIG}),
which is now gauge invariant.

To be an eigenstate of the magnetic field term $B^2$,
the vacuum of the system
is the fluctuation of the electric-field-line loops,
or a superposition of all loop configurations \cite{Levin2005PhysRevB}.
This enables the calculation of topological entanglement entropy.

Here we can identify the crucial chain of logic:
the gauge symmetry chosen determines
the Gauss' laws;
the gauge operators
are those objects (dipoles) obeying Gauss' law globally but not locally;
they  can be used to construct the  magnetic field that respect Gauss' law in 
any local region (minimal loops);
the magnetic  field determine the 
configuration of electric field lines at low energy (all loop configurations),
which then determine the entanglement structure of the system.

In Table.\ref{TAB_BIG}, the above logical chain is shown on the second row.

\begin{table*}[t]
\includegraphics[width=0.7\textwidth ]{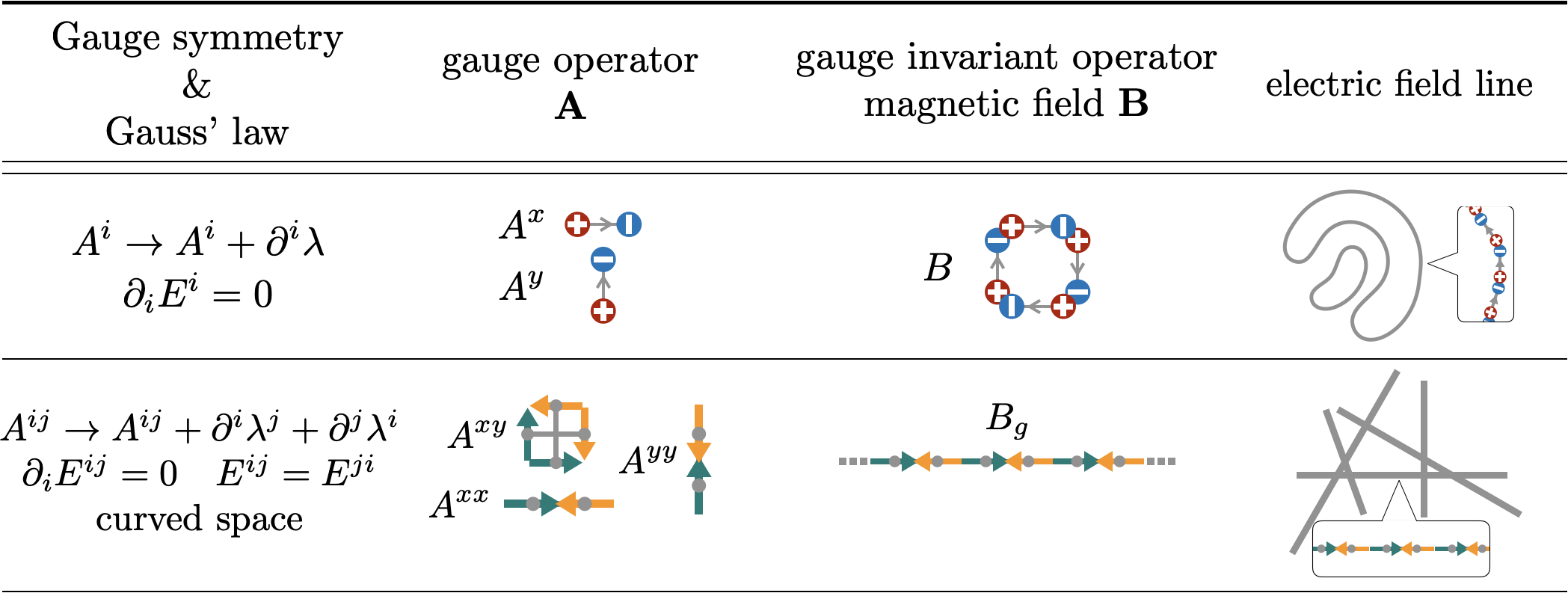}
\caption{From gauge symmetry to the entanglement structure.
This table demonstrates the logical chain leading  from
the gauge symmetry to the configurations of electric field lines
as the dynamical variables.
The second row is for conventional U(1), where the electric field lines can be arbitrary loops.
The third row is for rank-2 U(1) in AdS space, where the electric field lines 
are on the geodesics extending to infinity.
\label{TAB_BIG}
}
\end{table*}

\section{Entanglement structure of R2-U1 in AdS space: geodesic string condensation}

Now let us examine the entanglement structure 
of R2-U1 in the 2-dimensional AdS space following the same mechanism.
We will see that instead of string-net condensation,
the picture will  be \textit{``geodesic string condensation.''}
That is, 
the strings of electric fields 
travel along geodesics only,
and their superposition as the vacuum determines the entanglement structure.
The facts that the charge is a vector, and space is curved,
play crucial roles in determining the entanglement structure.

Like in the previous section,
the gauge symmetry and Gauss' laws
determine the effects of gauge operators
in terms of creating vector charge multipoles.
They are listed in Table.~\ref{TAB_BIG}.
The diagonal terms $A^{xx},\ A^{yy}$,
or in general $A^{ij}s_is_j$ for direction $\hat{\bm{s}}$  is to move
a vector charge along the direction it points.
The off-diagonal term $A^{xy}$ creates
a vector charge multipole with vanishing $\int \bm{\rho}$ and $\int \bm{\rho}\times  \bm{x} $.

The dynamics, or magnetic fields, however, are very different
in the curved space.
It has been carefully studied by Slagle et al. in Ref.~\cite{Slagle2018}.
When a vector charge is parallel transported
around a finite region back to its starting point,
it will in general be different from the original vector 
due to the spatial curvature.
So such parallel transport over a closed loop is energy-costly.

\begin{figure}[h]
	\centering
	\includegraphics[width=0.3\columnwidth]{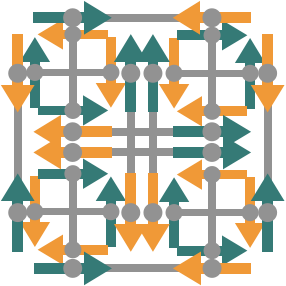}
	\caption{
The operator $B $ of rank-2 U(1) theory  in the flat space (Eq.~\eqref{EQN_R2U1_B}). 
It involves multiple   $A^{xx},\ A^{yy}$ and $A^{xy}$ operators. 
It acts on a finite-area plaquette in the system,
and does not survive the spatial curvature.
	} 
	\label{Fig_B_flat}
\end{figure}

Consequently, the local dynamics of ${B}$ (Eq.~\eqref{EQN_R2U1_B}) is forbidden.
The pictorial  intuition is that the dynamics of ${B}$ as illustrated by
Fig.~\ref{Fig_B_flat}
always happen over a finite-sized plaquette in the system.
In flat lattice, such combination of $A^{ij}$ operators 
does not violate the Gauss' laws (Eqs.~(\ref{EQN_R2U1_G1},\ref{EQN_R2U1_G1})) in any microscopic region.
But in curved space it is not true anymore.

To convince ourselves,
we find that
${B}$ (Eq.~\eqref{EQN_R2U1_B}) is not gauge invariant  
in the presence of curvature.
Promoting $\bm{\nabla}$ to the covariant derivative,
we have
\[
B \rightarrow B - Rg^{aj} \nabla_a \lambda_j		,
\]
under gauge transformation,
where $R$ is the Ricci scalar \cite{Slagle2018}.
In fact, higher order local terms up to $\nabla_a\nabla_b\nabla_c\nabla_d{\lambda}^e$
was systematically explored but no gauge invariant $B$ term was found in Ref.~\cite{Slagle2018}.

So what are the dynamics allowed in the AdS space?
We note that, 
the difficulty is rooted in parallel transporting 
a vector charge around a loop.
To avoid this,
we have to  consider instead parallel transporting
the charge on a geodesic, extending from
one infinity to the other.

In the lattice model,
it has the following picture:
A vector charge, for example, $\bm{\rho} = (\rho^x,0)$,
can be moved along $x-$direction by acting $A^{xx}$
operators on the path.
To make sure that any local region respects Gauss' laws,
however,
such line-operation has to extend to infinity in both directions.

In the field theory,
for a given geodesic $g$ with unit vector $\hat{\bm{s}}$ along it,
this operation is the dynamics
\[
B_g = \int_g \mathop{ds} A^{ij}\hat{s}_{i}\hat{s}_{j}.
\]
The fact that locally no Gauss' laws are broken is reflected
by its invariance under gauge transformation
\[
\begin{split}
B_g \rightarrow & B_g + \int_g\mathop{ds}  (\nabla^i\lambda^j + \nabla^i\lambda^j)\hat{s}_{i}\hat{s}_{j} \\
& =  B_g +2 (\bm{\lambda}\cdot\hat{\bm{s}}) \bigg\rvert^{\inf}_{-\inf},
\end{split}
\]
where the second term vanishes assuming vanishing gauge transformation at infinity.

We can thus write down the theory as 
\[
H_\text{R2-U1-AdS} = \int \mathop{dv} U E^{ij}E_{ij} +  \sum_{g\in \text{all geodesics}} t_g B_g^{\ 2}
\]

Such non-local dynamics are normally
unfavored in many disciplines of physics.
However, they are the ones 
stable in the presence of spatial curvature.
Let us bear with them,
and examine the corresponding of entanglement structure.

With such dynamics on geodesics,
a drastic change happens for the electric field lines.
In AdS space, instead of forming loops,
they travel along geodesics from one boundary point to another.
The vacuum is then a superposition of all possible
geodesic electric field line configurations.
We name this the ``geodesic string condensation.'' 
As a result, 
the entanglement structure 
for each geodesic string
is that
the two boundary points are entangled by the corresponding geodesic dynamics. 
As the $B_g$ distribute in AdS space homogeneously and isotropically,
we have exactly the continuous bit-thread picture we speculated at 
the beginning of this work.
Upon lattice discretization, and also assigning $\bm{E}$ discrete/continuous values,
one can obtain  toy models of the same universal picture but different in details,
including the perfect tensor-networks and the hyperbolic fracton models.

\section{Discussion}

In this work we obtained a very pictorial,
intuitive understanding of 
the ``leading order'' entanglement structure
of holography, and the mechanism  generating it.
The ``leading order'' entanglement structure
is a web of evenly distributed bit-threads,
and we noted that several holographic toy models belong to this picture.
We reason that,
taking the linearized diffeomorphism 
as the gauge symmetry,
the corresponding symmetric tensor gauge theory
gives rise to this picture by geodesic string condensation.
Retrospectively, 
it is sensible that  a theory mimicking gravity at first order
has  the entanglement structure of gravity
also at first order.

Many questions follow.
One exciting question to ask is that,
can we understand
the finer entanglement structure (some are listed in Table.~\ref{TAB_compare})
in a similar way? 
For example,
the random tensor-networks proposed by Yang et al. \cite{Yang2016,Hayden2016,Qi2018} 
satisfy the RT-formula for arbitrary  boundary subregion.
What   modification of the  geodesic string condensation  picture is needed 
to capture such properties?
Another question is how to introduce non-flat entanglement spectrum to match 
the $n-$th R\'{e}nyi entropy.
These projects will be very useful for us to gain improved intuition 
of the entanglement structure in holography.
The argument presented here is 
based on a chain of reasoning at a qualitative level.
It would deepen our understanding to 
re-derive these results through
more explicit calculations of the
entanglement entropy for the R2-U1, or its different variations,
on flat and AdS space.
It is also intriguing to know if 
the physics demonstrated in the work is connected to
other aspects of gravity, including 
its ground state degeneracy \cite{Hawking2016PhysRevLett},
and its relation to topological order \cite{Rasmussen2018PhysRevB}.

We hope our work will provide new, useful insight in
both understanding fracton states of matter and quantum gravity.

\section*{Acknowledgment}
We thank  Nic Shannon, Sugawara Hirotaka, Yasha Neiman, 
Xiao-Liang Qi, Kevin Slagle, Michael Pretko
for helpful discussions.
In particular we thank 
Nic Shannon, Sugawara Hirotaka and Yasha Neiman, 
for a careful reading of the manuscript.
HY is supported by 
the Theory of Quantum Matter Unit at Okinawa Institute of Science and Technology,
and
the Japan Society for the Promotion of Science (JSPS)
Research Fellowships for Young Scientists.

\bibliography{reference}

\end{document}